\documentclass[12pt]{article}
\usepackage{graphicx}
\begin{document}

\catcode`@=11
\long\def\@caption#1[#2]#3{\par\addcontentsline{\csname
  ext@#1\endcsname}{#1}{\protect\numberline{\csname
  the#1\endcsname}{\ignorespaces #2}}\begingroup
    \small
    \@parboxrestore
    \@makecaption{\csname fnum@#1\endcsname}{\ignorespaces #3}\par
  \endgroup}
\catcode`@=12
\newcommand{\newc}{\newcommand}
\newc{\lat}{{\ell at}}
\newc{\one}{{\bf 1}}
\newc{\mgut}{M_{\rm GUT}}
\newc{\mzero}{m_0}
\newc{\mhalf}{M_{1/2}}
\newc{\five}{{\bf 5}}
\newc{\fivebar}{{\bf\bar 5}}
\newc{\ten}{{\bf 10}}
\newc{\tenbar}{{\bf\bar{10}}}
\newc{\sixteen}{{\bf 16}}
\newc{\sixteenbar}{{\bf\bar{16}}}
\newc{\gsim}{\lower.7ex\hbox{$\;\stackrel{\textstyle>}{\sim}\;$}}
\newc{\lsim}{\lower.7ex\hbox{$\;\stackrel{\textstyle<}{\sim}\;$}}
\newc{\gev}{\,{\rm GeV}}
\newc{\mev}{\,{\rm MeV}}
\newc{\ev}{\,{\rm eV}}
\newc{\kev}{\,{\rm keV}}
\newc{\tev}{\,{\rm TeV}}
\newc{\mz}{m_Z}
\newc{\mw}{m_W}
\newc{\mpl}{M_{Pl}}
\newc{\mh}{m_h}
\newc{\mA}{m_A}
\newc{\mhpm}{m_{H^\pm}}
\newc{\hpm}{H^\pm}
\newc{\tr}{\mbox{Tr}}
\def\sfrac#1#2{{\textstyle\frac{#1}{#2}}}
\newc{\chifc}{\chi_{{}_{\!F\!C}}}
\newc\order{{\cal O}}
\newc\CO{\order}
\newc\CL{{\cal L}}
\newc\CY{{\cal Y}}
\newc\CH{{\cal H}}
\newc\CM{{\cal M}}
\newc\CF{{\cal F}}
\newc\CD{{\cal D}}
\newc\CN{{\cal N}}
\newc{\eps}{\epsilon}
\newc{\re}{\mbox{Re}\,}
\newc{\im}{\mbox{Im}\,}
\newc{\invpb}{\,\mbox{pb}^{-1}}
\newc{\invfb}{\,\mbox{fb}^{-1}}
\newc{\yddiag}{{\bf D}}
\newc{\yddiagd}{{\bf D^\dagger}}
\newc{\yudiag}{{\bf U}}
\newc{\yudiagd}{{\bf U^\dagger}}
\newc{\yd}{{\bf Y_D}}
\newc{\ydd}{{\bf Y_D^\dagger}}
\newc{\yu}{{\bf Y_U}}
\newc{\yud}{{\bf Y_U^\dagger}}
\newc{\ckm}{{\bf V}}
\newc{\ckmd}{{\bf V^\dagger}}
\newc{\ckmz}{{\bf V^0}}
\newc{\ckmzd}{{\bf V^{0\dagger}}}
\newc{\X}{{\bf X}}
\newc{\bbbar}{B^0-\bar B^0}
\def\bra#1{\left\langle #1 \right|}
\def\ket#1{\left| #1 \right\rangle}
\newc{\sgn}{\mbox{sgn}\,}
\newc{\m}{{\bf m}}
\newc{\msusy}{M_{\rm SUSY}}
\newc{\munif}{M_{\rm unif}}
\newc{\slepton}{{\tilde\ell}}
\newc{\Slepton}{{\tilde L}}
\newc{\sneutrino}{{\tilde\nu}}
\newc{\selectron}{{\tilde e}}
\newc{\stau}{{\tilde\tau}}
\newc{\vbb}{\beta\beta 0\nu}

\hyphenation{higg-strah-lung}
%
%
\def\NPB#1#2#3{Nucl. Phys. {\bf B#1} (19#2) #3}
\def\PLB#1#2#3{Phys. Lett. {\bf B#1} (19#2) #3}
\def\PLBold#1#2#3{Phys. Lett. {\bf#1B} (19#2) #3}
\def\PRD#1#2#3{Phys. Rev. {\bf D#1} (19#2) #3}
\def\PRL#1#2#3{Phys. Rev. Lett. {\bf#1} (19#2) #3}
\def\PRT#1#2#3{Phys. Rep. {\bf#1} (19#2) #3}
\def\ARAA#1#2#3{Ann. Rev. Astron. Astrophys. {\bf#1} (19#2) #3}
\def\ARNP#1#2#3{Ann. Rev. Nucl. Part. Sci. {\bf#1} (19#2) #3}
\def\MPL#1#2#3{Mod. Phys. Lett. {\bf #1} (19#2) #3}
\def\ZPC#1#2#3{Zeit. f\"ur Physik {\bf C#1} (19#2) #3}
\def\APJ#1#2#3{Ap. J. {\bf #1} (19#2) #3}
\def\AP#1#2#3{{Ann. Phys. } {\bf #1} (19#2) #3}
\def\RMP#1#2#3{{Rev. Mod. Phys. } {\bf #1} (19#2) #3}
\def\CMP#1#2#3{{Comm. Math. Phys. } {\bf #1} (19#2) #3}
\relax
%
%
%
\def\beq{\begin{equation}}
\def\eeq{\end{equation}}
\def\bea{\begin{eqnarray}}
\def\eea{\end{eqnarray}}
%
%
%
\newc{\ie}{{\it i.e.}}          \newc{\etal}{{\it et al.}}
\newc{\eg}{{\it e.g.}}          \newc{\etc}{{\it etc.}}
\newc{\cf}{{\it c.f.}}
\def\smuon{{\tilde\mu}}
\def\neut{{\tilde N}}
\def\char{{\tilde C}}
\def\bino{{\tilde B}}
\def\wino{{\tilde W}}
\def\higgsino{{\tilde H}}
\def\sneut{{\tilde\nu}}
%
%
%
%
\def\slash#1{\rlap{$#1$}/} 
\def\Dsl{\,\raise.15ex\hbox{/}\mkern-13.5mu D} 
\def\delsl{\raise.15ex\hbox{/}\kern-.57em\partial}
\def\Ksl{\hbox{/\kern-.6000em\rm K}}
\def\Asl{\hbox{/\kern-.6500em \rm A}}
\def\Qsl{\hbox{/\kern-.6000em\rm Q}}
\def\gradsl{\hbox{/\kern-.6500em$\nabla$}}
%
%
%
\def\bar#1{\overline{#1}}
\def\vev#1{\left\langle #1 \right\rangle}
%

\begin{titlepage}
\begin{flushright}
\end{flushright}
\vskip 2cm
\begin{center}
{\large\bf
Constraining the Charged Higgs Mass in the MSSM: \\ A Low-Energy Approach}
\vskip 1cm
{\normalsize\bf
Brian Dudley and Christopher Kolda\\
\vskip 0.5cm
{\it Department of Physics, University of Notre Dame\\
Notre Dame, IN~~46556, USA}\\[0.1truecm]
}

\end{center}
\vskip .5cm

\begin{abstract}
We examine the current lower bound on the charged Higgs boson mass in the Minimal Supersymmetric Standard Model. By applying direct search constraints on the neutral Higgs bosons and other supersymmetric states, as well as a number of indirect constraints ($b\to s\gamma$, $B_s\to\mu\mu$, $B\to\tau\nu$, $B\to D\tau\nu$), we find that it is possible to push the charged Higgs boson mass as low as $140\gev$. We work in a completely low-energy approach with no assumptions about scalar mass unification, examining two of the most popular limits for neutral Higgs boson searches: the Max-Mixing and No-Mixing scenarios. While both scenarios allow light charged Higgs bosons, they do so for completely different ranges of $\tan\beta$. In either case, one expects light top squarks to accompany a light charged Higgs.
\end{abstract}

\end{titlepage}

\setcounter{footnote}{0}
\setcounter{page}{1}
\setcounter{section}{0}
\setcounter{subsection}{0}
\setcounter{subsubsection}{0}

With the initiation
of the LHC, physicists will soon be able to search for new physics at energy regimes
never before possible.  Among the possible discoveries that would signal the existence of physics beyond the Standard Model (SM), one that stands out (among several) would be the discovery of a charged Higgs boson. While physicists have become comfortable thinking of the SM as a theory with a single Higgs SU(2) doublet, there are simple extensions, trivially consistent with all available data, in which additional SU(2) singlets and/or doublets are added to the spectrum of the Higgs sector. One of the smoking guns for the models with additional doublets, called Two-Higgs Doublet Standard Models (2HDSMs), are physical, charged Higgs bosons.

\begin{figure}[tb]
\centering
\includegraphics[scale = 0.7]{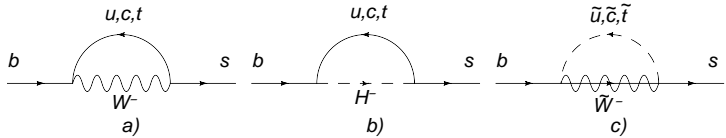}
\caption{Feynmann diagrams contributing to $b\to s\gamma$. Includes
(a)~$W^{\pm}$, (b)~$H^{\pm}$ and (c)~$\chi^{\pm}$ diagrams.
\label{bsgdiags}}
\end{figure}

If a charged Higgs ($\hpm$) exists, there are a number of channels, both direct and indirect, in which its presence could have a profound effect. But the most well-known (and currently most limiting) constraint on the $\hpm$ comes from the rare decay $b\to s\gamma$. Constructive interference in 2HDSMs between a $\hpm$-mediated loop diagram and the standard $W^\pm$-mediated diagram (see Fig.~\ref{bsgdiags}) leads to lower bounds on the charged Higgs mass of roughly $300\gev$~\cite{Hewett:1992is}.

However, it has long been known that $b\to s\gamma$ can have other, equally important contributions in more general models of new physics. In particular, within the Minimal Supersymmetric Standard Model (MSSM), chargino- and neutralino-mediated diagrams (Fig.~\ref{bsgdiags}(c)) can partially or completely cancel the charged Higgs contribution, eliminating the aforementioned bound~\cite{Barbieri:1993av}.
Given the possibility of at least some cancelation amongst the different contributions, it is almost taken for granted that supersymmetry (SUSY) can accommodate a lighter charged Higgs than the 2HDSMs.
However, the other light SUSY particles which help in the $b\to s\gamma$ cancelation also enhance other flavor changing processes, such as $B_s\to\mu\mu$. At the same time, a charged Higgs still mediates tree-level flavor-changing process such as $B\to\tau\nu$ and $B\to D\tau\nu$. All of these processes provide constraints of varying strength on both
the charged Higgs and other SUSY masses, which makes determining a direct theoretical mass bound on
the charged Higgs difficult to achieve.

There are additional constraints on both the charged Higgs and the spectrum of SUSY partners (``sparticles") which arise from the neutral Higgs sector. Within the MSSM, the charged Higgs is accompanied by three neutral Higgs bosons, the lightest of which is often SM-like, and for which a number of experiments have conducted searches. At tree level, the lightest Higgs ($h$) must be lighter than the $Z$; at higher order the mass can be raised as high as 125 to $135\gev$, depending on the details of the sparticle spectrum and the masses of the other Higgs states. These correlations can be quite complicated and, as we will see, can rules out many models with a light charged Higgs.

In this paper we examine all these varied constraints in order to calculate a true lower bound on the charged Higgs mass, an important question for phenomenologists at the start of the LHC era.
This question has been examined in recent literature~\cite{Domingo:2007dx,Barenboim:2007sk,Eriksson:2008cx},
and like these papers, we
include various $B$-physics and electroweak constraints.  However, we will try to maintain, in so far as possible, a completely ``bottom-up" analysis; that is, we make no assumption about unification at some high scale, which means we avoid the hidden correlations that usually arise in that class of models. We also will work within two limits that are most-often examined by phenomenologists and experimentalists looking for the light Higgs: the so-called Max-Mixing and No-Mixing scenarios, which we will define in the next section. In a sense, we trade some of the high-scale simplifications of Ref.~\cite{Domingo:2007dx,Barenboim:2007sk,Eriksson:2008cx} in favor of low-energy simplifications. Our final results, however, as in broad agreement with the earlier analyses, as we will explain.

The plan for this paper is as follows: We begin with three sections reviewing the major constraints, both direct and indirect, on the Higgs sector of the MSSM, and on the sparticle spectrum.
Following that discussion, we explain in $\S4$ how our analysis was performed, the results of which are discussed in detail in $\S5$. These results are compared to earlier results in the Conclusions ($\S6$).

\section{Direct Observables}

The first direct observable of importance is obviously the search
for the charged Higgs with the current experimental limit coming
from the process $e^+e^-\to H^+H^-$. The best searches so far only
place a lower limit on $m_{H^{\pm}}$ of $79\gev$, which is far below
indirect searches in 2HDMs.  However, the parameters in the Higgs
sector of SUSY are all strongly connected. At tree level, the
charged Higgs mass depends on the CP-odd Higgs, $A$, through the
relationship: \bea m_{H^\pm}^2=m_A^2+m_W^2. \eea Many other
relationships like this exist amongst the Higgs parameters and at
tree level the Higgs spectrum in SUSY is usually determined by two
parameters, commonly taken to be the mass of the CP-odd Higgs boson,
$m_A$ and $\tan\beta = v_u/v_d$. Through these relationships direct
limits on other Higgs particles impact the mass range available to
the charged Higgs.

Two direct limits of interest are those on $m_A$ and the lightest
CP-even Higgs mass, $m_h$.  The lightest CP-even Higgs, $h$, in
particular has been strongly sought after and in SUSY is quite
light, lighter than $m_Z$ at tree level. However, quantum
corrections play a large role, and can raise $m_h$ above the LEP
bound of $114\gev$. The primary quantum corrections of interest are
those generated by stop loops and yielding a light Higgs
mass~\cite{Carena:1995bx}: \bea m_h^2 & = &
m_Z^2\cos^22\beta\left(1-\frac{3m_t^2}{8v^2\pi^2}
\log\frac{\msusy^2}{m^2_t}\right)\nonumber\\
&&
+\frac{3m_t^4}{4v^2\pi^2}\left[\frac{X_t^2}{\msusy^2}\left(1-\frac{X_t^2}{12\msusy^2}
\right)+\log\frac{\msusy^2}{m^2_t}\right]. \label{HiggscorrectionEq}
\eea In the above $X_t = A_t -\mu\cot\beta$ and $\msusy$ is
understood to be the average mass of the top squarks.

The size of these corrections can vary greatly, and in many
phenomenological studies, the choice is reduced to two extremes,
these being the ``Max-Mixing" and ``No-Mixing" scenarios. The
Max-Mixing scenario occurs when $m_h$ is maximized by setting $X_t =
\sqrt{6}\msusy$, whereas the No-Mixing scenario sets $X_t = 0$. The
former case generally implies large $A$-terms while the latter
typically requires a heavier $\msusy$ in order to pass the LEP
bound.

The lower limit on the Higgs mass of $114\gev$ was obtained through
a search for the SM higgstrahlung process ($e^+e^-\to Zh$) at LEP.
SUSY effects can alter the bound significantly because of a
suppression in the coupling between the Z boson and the light Higgs
($g_{ZZh}$) which leads to a suppression of the cross section of the
form: \bea \sigma^{hZ}_{\mbox{\tiny{SUSY}}}& = &
\sin^2(\beta-\alpha)\times\sigma^{hZ}_{\mbox{\tiny{SM}}}. \eea In
the above, $\sigma^{hZ}_{\mbox{\tiny{SM}}}$ is the Standard Model
cross section given by: \bea
\sigma^{hZ}_{\mbox{\tiny{SM}}} & = &C_{EW} \lambda^{1/2}\left(\lambda+12sm_Z^2\right),\nonumber\\
\lambda & = & (s - m_h^2 - m_Z^2)^2 - 4m_h^2m_Z^2, \eea where, we
condense the pieces of the calculation not dependent on $m_h$ into a
constant $C_{EW}$. This constant depends on the center of mass
energy, $s$, and other electroweak parameters such as $m_Z$,
$\alpha$ and the weak angle, $\theta_W$.

LEP also considered the production of the CP-odd Higgs boson, $A$.
The cross section for $e^+e^-\to Z\to hA$ is suppressed by
$\cos^2(\beta-\alpha)$ which goes as: \bea \cos^2(\beta -\alpha) =
\frac{m_h^2\left(m_Z^2-m_h^2\right)} {m_A^2\left(m_H^2-m_h^2\right)}
\eea In the Higgs decoupling limit, where $m_A$ is large,
$\cos^2(\beta -\alpha)\to 0$ which leaves the higgstrahlung process
above SM-like. These two experiments are complementary and combine
for a robust constraint on the light CP-even Higgs.

We implemented the constraint from the SM Higgs search by requiring
the the higgstrahlung cross section for the SUSY Higgs, $h$,
satisfies the following bound: \bea
\sigma^{hZ}_{\mbox{\tiny SUSY}}\left(m_h;\sqrt{s}=209\gev\right)
\le \sigma^{hZ}_{\mbox{\tiny SM}}\left(m_h = 114, \sqrt{s} =
209\gev\right). \eea 
For the complementary process, $e^+e^-\to Z\to
hA$, the decay of the $A$ boson is unfortunately very model
dependent. Therefore we will enforce the LEP constraint on $m_A$ of
$91\gev$ though it is possible that a lighter $A$ could be possible.

Finally, LEP has also placed limits on other SUSY particles. For
instance, the lightest stop, $\tilde{t}_1$, is required to be
heavier than $95\gev$ and the lightest chargino, $\chi^{\pm}_1$, must
be heavier than $103\gev$. This last limit constrains the
$\left|\mu\right|$ parameter through the chargino mass matrix. While
other SUSY particles have direct limits imposed by LEP, such as the
lightest sbottom, $m_{\tilde{b}_1}$ and lightest neutralino,
$\chi^0_1$, they do not have a sizeable effect for the parameter
space we examine in this analysis  More details about these
constraints is given in section 4.

\section{B Physics Observables}

Over the last decade, the number of measured observables and their
precision has increased dramatically. These measurements have in
turn provided new, powerful constraints on beyond the SM physics and
SUSY in particular. In this analysis many different flavor-violating
processes are considered which are enhanced by the presence of a
light charged Higgs.  These include processes such as $b\to
s\gamma$, which currently places a very strong limit on
$m_{H^{\pm}}$ in 2HDMs. There is also $B_s\to\mu\mu$, which receives
corrections from SUSY which are an order of magnitude greater than
the SM contributions. We study these processes and other flavor
observables, such as the decays $B\to\tau\nu$ and $B\to D\tau\nu$
which are tree-level processes mediated by a charged Higgs.

\subsection{$b\to s\gamma$}

The calculation of $b\to s\gamma$ is generally broken into two parts
by way of the operator product expansion.  The short range physics,
which includes all the contributions from SUSY, is contained in the
calculation of the Wilson coefficients,${\cal C}_i$.  The
calculation of these coefficients is well documented in the
literature.~\cite{Barbieri:1993av,Bertolini:1990if} The $W^{\pm}$
and $H^{\pm}$ contributions, which are shown in
Fig.\ref{bsgdiags}a)-b), are as follows: \bea
{\cal C}_{W}& = & \frac{3}{2}\frac{m_t^2}{m_W^2}\mbox{F}_1\left(\frac{m_t^2}{m_W^2}\right),\nonumber\\
{\cal C}_{H^{\pm}}& = & \frac{1}{2}\frac{m_t^2}{m_{H^{\pm}}^2}
\left(\frac{1}{\tan^2\beta}\mbox{F}_1\left(\frac{m_t^2}{m_{H^{\pm}}^2}\right)+
\mbox{F}_2\left(\frac{m_t^2}{m_{H^{\pm}}^2}\right)\right), \eea
where $\mbox{F}_1$ and $\mbox{F}_2$ are kinematic loop functions
given in~\cite{Bertolini:1990if}.  These two contributions are the
only ones found in 2HDMs and they always constructively interfere.
This allows the current measurements on $b\to s\gamma$ to place
strict limits on the charged Higgs mass in such models, with current
bounds requiring the charged Higgs in 2HDMs to be heavier than
$\approx300\gev$.

Within SUSY, the charged Higgs and SM contributions also interfere
constructively, but now there are the chargino contributions, shown
in Fig.~\ref{bsgdiags}c), which depend on several different
particles masses and the chargino and stop mixing angles; these in
turn bring in a dependence on soft SUSY parameters such as $\mu$ and
$A_t$. The chargino contributions usually interfere destructively
with the SM, which weakens the limit on $m_{H^{\pm}}$ coming from
$b\to s\gamma$.

To determine the effect of this short range physics, the entire
calculation, including the long range QCD effects, must be
considered. Converting the Wilson coefficients into a rate for
$\mbox{Br}(B\to X_s\gamma)$ has been well
studied~\cite{Hurth:2003dk,Misiak:2006zs}. We use the NLL
calculation of Ref.~\cite{Hurth:2003dk}, which easily and
analytically incorporates new physics from the Wilson coefficients.
The largest source of error in this calculation comes from the charm
mass in the ratio $m_c/m_b$ and we use this large uncertainty to
tune the NLL calculation so that it reproduces the NNLO theoretical
value for the SM calculation of $\mbox{Br}(B\to X_s\gamma)$ of
$3.15\times 10^{-4}$ when the new physics is turned off.

This theoretical value is lower than the current experimental value
which is given by the Heavy Flavor Averaging
Group~\cite{Barberio:2008fa} to be: \bea \mbox{Br}(B\to
X_s\gamma)_{exp}&=& (3.55\pm0.24^{+0.09}_{-0.10}\pm0.03)\times
10^{-4}. \eea This allows some room for new physics, and leads to
the following $2\sigma$ confidence interval which we enforce on our
analysis: \bea 3.03\times10^{-4} < & \mbox{Br}(B\to
X_s\gamma)_{E_{\gamma}>1.6\mbox{GeV}}& < 4.06\times 10^{-4}. \eea

\subsection{$B_s\to\mu\mu$}

\begin{figure}[htb]
\centering
\includegraphics[scale = 1.2]{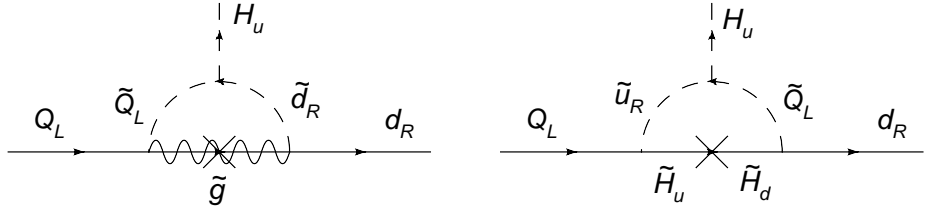}
\caption{Higgs Penguin diagrams contributing to
$B_s\to\mu\mu$.\label{higgspenguins}}
\end{figure}

Another probe of SUSY, which is particularly powerful at large
$\tan\beta$, is the decay $B_s\to\mu\mu$. This process is mediated
by Higgs penguin diagrams which create an effective coupling between
$H_u$ and the down type quarks: $H_u\bar{d}_R\Delta_uq_L$. This
effective coupling picks up contributions from two sources in SUSY,
a gluino loop which is flavor preserving and a chargino loop which
is flavor violating as shown in Fig.~\ref{higgspenguins}. The size
of this effective coupling is given
by~\cite{Babu:1999hn,Buras:2002wq}: \bea (\Delta_u)^{JI} &=&
y_{d_J}\left(\epsilon_0\delta^{IJ}+\epsilon_Yy_t^2V^{3J*}V^{3I}\right),
\eea where $I,J=1...3$ designate the quark generation, and \bea
\epsilon_0& = & -\frac{2\alpha_s}{3\pi}\frac{\mu}{m_{\tilde{g}}}H_2(x^{Q/\tilde{g}},x^{D/\tilde{g}}),\nonumber\\
\epsilon_Y& = &
\frac{1}{16\pi^2}\frac{A_t}{\mu}H_2(x^{Q/\mu},x^{U/\mu}). \eea Here
$x^{Q/\mu}=m_Q^2/\mu^2$ and the other $x$'s are similarly defined
and $H_2$ is a kinematic function defined to be: \bea
H_2(x,y)&=&\frac{x\ln x}{(1-x)(x-y)}+\frac{y\ln y}{(1-y)(y-x)}. \eea

Two of the prominent features of this decay are its strong
dependence on $\tan\beta$ and $m_A$, which enter the branching ratio
with six and four powers respectively. In the large $\tan\beta$
limit, where the SUSY contribution is clearly important, the
branching ratio can be formulated as~\cite{Buras:2002wq}: \bea
\mbox{Br}(B_s\to\mu\mu)& = &
3.5\times10^{-5}\left[\frac{\tan\beta}{50}\right]^6
\left[\frac{\tau_{B_s}}{1.5 ps}\right]\left[\frac{F_{B_s}}{230\mbox{
MeV}}\right]^2
\left[\frac{\left|V_{ts}\right|}{0.040}\right]^2\nonumber\\
&&\times
\frac{m_t^4}{m_A^4}\frac{(16\pi^2)^2\epsilon_Y^2}{(1+\tilde{\epsilon}_3\tan\beta)^2
(1+\epsilon_0\tan\beta)^2}. \eea In the above, \bea
\tilde{\epsilon}_3 & = & \epsilon_0+\epsilon_Yy_t^2.\eea

This large $\tan\beta$ enhancement allows the SUSY contribution to
this decay to dominate over the SM prediction, which is highly
suppressed: $\mbox{Br}(B_s\to\mu\mu)_{\mbox{\tiny SM}} =
(3.2\pm0.5)\times 10^{-9}$. Such a large possible contribution from
new physics has driven the search for this decay. Currently, the
strongest experimental limit comes from CDF~\cite{:2007kv}: \bea
\mbox{Br}(B_s\to\mu\mu)_{\mbox{\tiny exp}} &< &5.8\times10^{-8}.
\eea

\subsection{$B\to\tau\nu$}

The process $b\to s\gamma$, though it is well-studied and provides a
strong constraint, only depends on the charged Higgs through loops.
This allows other SUSY particles to interfere with the effect of the
charged Higgs contribution as described above. However, SUSY models
with a charged Higgs also generate tree-level flavor violation and
without the loop factors, ${\cal O}(1)$ corrections to several
processes can be produced. One process of interest is $B\to\tau\nu$
which is generated by a simple tree-level exchange of a $H^{\pm}$.
The contribution of the charged Higgs to $B\to\tau\nu$ has been
studied in many sources~\cite{Hou:1992sy,Akeroyd:2003zr} and the
decay rate is given by: \bea
\Gamma(B^+\to\tau\nu)&=&\frac{G_F^2m_Bm_{\tau}^2f_B^2}{8\pi}\left|V_{ub}\right|^2
\left(1-\frac{m_{\tau}^2}{m_B^2}\right)^2\times r_H, \eea where
$r_H$ contains the $m_{H^{\pm}}$ dependence: \bea r_H =
\left(1-\frac{\tan^2\beta}{1+\epsilon_0\tan\beta}\frac{m_B^2}{m_{H^{\pm}}^2}\right)^2.
\eea In the equation above, the $\epsilon_0$ is the same Higgs penguin
corrections discussed in the $B_s\to\mu\mu$ section.

The current experimental limits for $B\to\tau\nu$ is often expressed
as a limit on the ratio of the experimental value and the SM
theoretical calculation~\cite{Barberio:2008fa}: \bea
R_{B\to\tau\nu}=\frac{\mbox{Br}(B\to\tau\nu)_{\mbox{exp}}}{\mbox{Br}(B\to\tau\nu)_{\mbox{SM}}}&
=& 1.28\pm0.38. \eea We turn this into a $2\sigma$ confidence
interval on $R_{B\to\tau\nu}$: \bea 0.52<&R_{B\to\tau\nu}<2.04 \eea

However, while $B\to\tau\nu$ can provide a strong experimental
constraint, as found in~\cite{Barenboim:2007sk}, its calculation is
plagued by large theoretical uncertainties; neither the decay
constant, $f_B$, nor the CKM element $V_{ub}$ are well measured.
$V_{ub}$ in particular is poorly understood with a large range of
experimental measurements which depend on whether the measurement is
from the inclusive or exclusive decays. In fact, several of the most
precise measurements of $V_{ub}$ obtain very different
results~\cite{Barberio:2008fa,Bona:2006ah}. This presents a problem when trying to
interpret the effect of this constraint on our models and we will discuss it further in
section $5$.

\subsection{$B\to D\tau\nu$}

While $B\to\tau\nu$ has a large theoretical uncertainty due to the
error in $V_{ub}$, $B\to D\tau\nu$ depends on the more accurately
measured $V_{cb}$. Unfortunately,  the presence of multiple
neutrinos in the final state makes the detection of $B\to D\tau\nu$
more of an experimental challenge. But, with growing certainty in
the measurement, it can possibly provide comparable power to
constrain new physics models.

In order to limit the dependence on hadronic parameters, it is more
useful to calculate a ratio of $B\to Dl\nu$ decays. We follow the
results of Refs.~\cite{Kamenik:2008tj,Nierste:2008qe} to obtain the
ratio: \bea \frac{\mbox{Br}(B\to D\tau\nu)}{\mbox{Br}(B\to
De\nu)}&=& (0.28\pm 0.02)\times\left(1+ 1.38\,Re(C_{NP}) +
0.88\left|C_{NP}\right|^2\right). \eea In the above, $C_{NP}$
contains all the contributions that come from new physics.
Neglecting all but the tree-level charged Higgs contributions, one
finds: \bea C_{NP} & = & -
\frac{m_bm_{\tau}}{m^2_{H^{\pm}}}\frac{\tan^2\beta}{1+\epsilon_0\tan\beta}.
\eea

We compare our calculation to the measurement by
BaBar~\cite{Aubert:2007dsa}: \bea \frac{\mbox{Br}(B\to
D\tau\nu)}{\mbox{Br}(B\to De\nu)} &=& 0.416\pm 0.117\pm 0.052. \eea
Which leads to a $2\sigma$ confidence interval of:
\beq
0.151<\frac{\mbox{Br}(B\to D\tau\nu)}{\mbox{Br}(B\to De\nu)}<0.681.
\eeq

\section{Other Observables}

The observables listed in the previous sections are the dominant constraints on the charged Higgs mass, and they are also the most model independent. There are however a number of other constraints which are traditionally applied to the parameter space of the MSSM, some of which we use and others we choose to ignore, for reasons we outline now.

So far, we have applied nothing in the way of constraints on the slepton sector. Though this sector has a
rather rich phenomenology, constraints arising here
generally depend on slepton masses and mixings, and on complicated, highly model-dependent decay rates among the
neutralinos and sleptons.  Such observables however, exhibit very weak, or no, direct
dependence on the charged Higgs mass. If we were to implement some
unification constraints on our models, such as is often done in the
literature (\eg, a unified scalar mass, $M_0$), the sleptons become linked to
the squark sector and therefore in turn to the Higgs sector. Enforcing such
a constraint on the mass spectrum will of necessity result in tighter limits on $\mhpm$
which are higher than in more general models. Since our goal is to find something close to an
absolute lower bound on $\mhpm$, we will not take this approach. However there are still a few,
generally useful constraints that we can learn from physics in the (s)leptonic sector.

The main such constraint arises from the measurement of the muon anomalous magnetic moment, denoted $a_\mu$. Both the theoretical and experimental values of $a_\mu$ have been extracted to extraordinarily high precision, with a small (roughly $3\sigma$) discrepancy found between experiment and the SM.
The current difference between experiment and the SM
calculation is~\cite{Bennett:2006fi,Miller:2007kk}:
$$
a_{\mu}^{\mbox{\scriptsize exp}} - a_\mu^{\mbox{\scriptsize SM}}  =   (29.5 \pm 8.8)\times 10^{-10}
$$
If the spectrum of SUSY is at all light, then it will contribute to this
anomaly, but with a sign that depends almost solely on the sign of
the $\mu$-parameter~\cite{Feroz:2008wr}. Given the sign of the discrepancy,
a positive sign for $\mu$ is favored (using here the Les Houches conventions~\cite{Allanach:2008qq}). Therefore
we will present our results under the assumption of $\mu>0$, leaving the case for $\mu<0$ for our comments
at the end of the next section.

But since the sleptons play little other direct role in
constraining $\mhpm$, we will push all slepton masses to $10\tev$ so
that they don't inadvertently create correlations which are not generic to the MSSM.

Another possible constraint comes from astrophysics. Observations~\cite{Komatsu:2008hk} have shown that
there is an abundance of dark matter in the universe which cannot
be explained within the SM.  Low energy SUSY models generally introduce a
new symmetry, called $R$-parity, in order to avoid fast proton decay. In the process, $R$-parity forces the lightest SUSY particle (LSP) to be stable, making it a candidate for the universe's dark matter.
Ideally there are two sets of constraints on the LSP. First, it must be electrically neutral so that the matter is truly ``dark", and second it must produce the correct density of dark matter as observed by astronomers. We will use the first constraint (though, as we will see, it does very little for us), but ignore the second constraint, since constraints on the calculated relic abundance once again produce a great deal of model dependence in our results.

\section{Analysis}

The goal of this analysis is to find a lower bound on the charged Higgs mass within the MSSM, but without enforcing ``unnatural" constraints on the model, such as unification of all scalar masses. As we will see shortly, the analysis here is not entirely general, but captures most of the physics that a truly general analysis would uncover.

One way in which we avoid being completely general is in our treatment of the radiative corrections to the lightest neutral Higgs boson. Studies of the Higgs sector phenomenology usually consider two particular limits for the lightest Higgs mass, called the ``Maximal Mixing" and ``No Mixing" scenarios. The names refer to the size of the top squark mixing corrections that are calculated at the SUSY decoupling scale, as discussed earlier. All of our calculations are done within one or the other of these scenarios.

Our basic procedure is to perform a random scan over the low-energy parameters of the MSSM, enforcing all constraints and then finding the lower bound on $\mhpm$ consistent with those constraints. For the plots that follow, we have shown only 10,000 points, though we have studied the parameter space with many more, particularly near the boundaries where a transition between constraints arises.

The parameters in the scan are entirely weak-scale parameters; no unification or running is done in the analysis. But these parameters are generally not equal to physical masses, and we include F- and D-term contributions, left-right mixing, and leading one-loop corrections in calculating physical mass eigenvalues. In the next few paragraphs we summarize the constraints that we place on the input parameters in order to define our model space.

We begin by setting the slepton masses to be very heavy -- $10\tev$ -- and $\mu$ positive. As we mentioned in the previous section, the observables which depend crucially on
slepton masses, such as the dark matter relic density or $a_\mu$,
have very little correlation with the charged Higgs mass.  By setting
the slepton masses high, and by choosing that our models have a positive sign for $\mu$, we include
the indications given by these constraints without ruling out otherwise ``good" models unnecessarily. Another way to say this is as follows: if a model point is acceptable in every way except that it generates too much/too little dark matter, or falls outside the experimentally allowed range for $a_\mu$, we can probably fix that problem by shifting the slepton masses, without affecting the charged Higgs mass in any way. Thus we choose to essentially decouple the sleptons from the start.

In the squark sector we set the masses of the first two generations to $1\tev$, for reasons that echo those for the sleptons. However the third generation plays a key role in the charged Higgs bound, coming into both the radiative corrections to the Higgs masses themselves, and the chargino-stop diagrams that contribute to $b\to s\gamma$. Thus we allow the the third generation soft squark mass parameters to run within the range of
$500\gev$ to $1\tev$. To make the analysis simpler, we assume that the soft mass parameters for each of the third generation squark states (the left-handed stop-sbottom doublet, and the right-handed stop and sbottom singlets) all share a common soft mass, $m_{\tilde{q}_3}$. From the soft masses, the physical masses are calculated by adding the ``threshold corrections": F- and D-term pieces, as well as left-right mixing. For the left-right mixing, we will need to choose values for the $A$ and $\mu$-terms, which we will discuss shortly. Some loss of generality is unavoidable by our choice of a universal third generation squark mass. Even so, after including threshold corrections to the squark masses, the squarks can become as light as $300\gev$, where they would start to run into Tevatron constraints.

As for the left-right mixing, we set all $A$-terms, except $A_t$, to be zero. (We assume that trilinear soft breaking terms are proportional to their corresponding Yukawa matrices, which justifies this simplification.) Instead of varying $A_t$ as another free parameter, it instead takes on two
extreme values that have been used in many phenomenological studies
of SUSY. The first is the Max-Mixing scenario in which
$X_t=A_t-\mu\cot\beta$ is given a value such that the correction to
the light Higgs mass in Eq.~(\ref{HiggscorrectionEq}) is maximized.
Once $\mu$ and $\tan\beta$ are chosen in a model, $A_t$ is chosen
such that $X_t = \sqrt{6} m_{\tilde{q}_3}$.  The second scenario is
the No-Mixing scenario in which $A_t = \mu\cot\beta$ and $X_t$ is
identically zero.

In the gaugino sector, we set the bino mass, $M_1$, and the gluino mass, $M_3$, to fixed values, while
allowing the wino mass, $M_2$, to vary between 100 and $500\gev$. Varying $M_2$ is important because, along with
the $\mu$ parameter it directly enters the diagrams for $b\to s\gamma$
through the chargino masses. $M_1$, however, does not need to vary,
as the only constraint that depends heavily on its value is the LSP dark matter
constraint. With the sneutrino masses raised to $10$ TeV, the
only remaining candidate for dark matter is the lightest neutralino,
whose mass is tied strongly to $M_1$.  For this analysis, $M_1 =
60\mbox{ GeV}$, a value which both satisfies the LEP bound on
$\chi^0_1$, and provides for $\chi^0_1$ to be the LSP. (There is a small dependence in $b\to s\gamma$ on $M_1$ through the neutralino diagrams, but this effect is far too small to affect our results.)

The gluino mass, $M_3$, has no leading contributions to any of the observables we are
considering, but it does have one sub-leading effect. Because the non-holomorphic corrections
to the down-quark masses depend on $M_3$ (through the parameter $\epsilon_0$), there is a second
order dependence in both $B\to\tau\nu$ and $B\to D\tau\nu$. Even $B_s\to\mu\mu$, which depends
heavily on the non-holomorphic terms, is primarily dependent only on $\epsilon_Y$ (which has no $M_3$ dependence),
and only receives a small correction via the $M_3$-dependent $\epsilon_0$ term. For these
reasons, $M_3$ in this study is set to a fixed value of $1\tev$, at which most of its effects
decouple.~\footnote{We note in passing that while the observables considered here have little to no dependence
on $M_3$ or $\epsilon_0$, there appears to be a rather strong dependence on these parameters in the
discovery potential at the LHC. See Ref.~\cite{Hashemi:2008ma} for details.}

The $\mu$-parameter is also varied: $100\gev\leq \mu\leq 1\tev$. The value of $\mu$ is assumed to remain positive for all of the results shown in the next section (consistent with $a_\mu$), but we have tested the stability of our results for negative $\mu$ and find no new regions of parameter space appear. The lower bound of $100\gev$ is set by the LEP limit on chargino masses.

We treat the Higgs sector much like the others -- we vary over a set of input parameters and
check to see if the resulting set of inputs is consistent with all of our constraints. Normally
one parametrizes the Higgs sector with two parameters, $\tan\beta$ and $m_A$.  But for the purposes of this
paper, it is more convenient to replace $m_A$ with $\mhpm$ which
is allowed to vary between 100 and $300\gev$. The upper bound is set because above $300\gev$ it is rather
trivial to find models consistent with the $b\to s\gamma$ constraint, and so we have no interest there.
The parameter $\tan\beta$ is varied from $1\leq \tan\beta\leq 70$. Allowing $\tan\beta$ to push above the more usual limits of 50--60 will have some effect in the No-Mixing scenario.

It is also important to note while our calculation of the Higgs spectrum includes many of the
first order corrections, however, there are higher order corrections which can
shift the light Higgs mass by $3-5\gev$~\cite{Allanach:2004rh}. To account for this
uncertainty and guarantee models are not incorrectly thrown out, $3\gev$ is added to the masses of $A$ and $h$ before
enforcing the calculating the observables and applying their constraints.

The range over which the various parameters have been varied is summarized in Table~\ref{ParameterTable}.
\begin{table}[tb]
\centering
\begin{tabular}{|c|c|}
\hline
$\tan\beta$&$1 - 70$\\
\hline
$m_{H^{\pm}}$&$100 - 300\mbox{ GeV}$\\
\hline
$\mu$&$100 - 1000\mbox{ GeV}$\\
\hline
$M_1$&$ 60 \mbox{ GeV}$\\
\hline
$M_2$&$ 100 - 500 \mbox{ GeV}$\\
\hline
$M_3$&$1000\mbox{ GeV}$\\
\hline
$m_{\tilde{q}_{1,2}}$&$1\mbox{ TeV}$\\
\hline
$m_{\tilde{q}_3}$&$500 - 1000\mbox{ GeV}$\\
\hline
$m_{\tilde{\ell}}$&$10\mbox{ TeV}$\\
\hline
\end{tabular}
\caption{Range of parameters in MSSM scan. \label{ParameterTable}}
\end{table}

Once a set of parameters is chosen in either of the Max-Mixing/No-Mixing scenarios, a physical spectrum is generated. In order to take include the corrections necessary in both the squark and Higgs sector, we started from the publicly available CPsuperH code~\cite{Lee:2003nta}, altered to our uses. Once a physical spectrum and its accompanying parameters (such as the angle $\alpha$ in the neutral Higgs sector), each observable was calculated and compared to its $2\sigma$ confidence bounds. The results of this analysis, in terms of which points passed all the constraints and how each constraint sliced the parameter space, is discussed below.

\section{Results}

\begin{figure}[t]
\centering
\includegraphics[scale = .50]{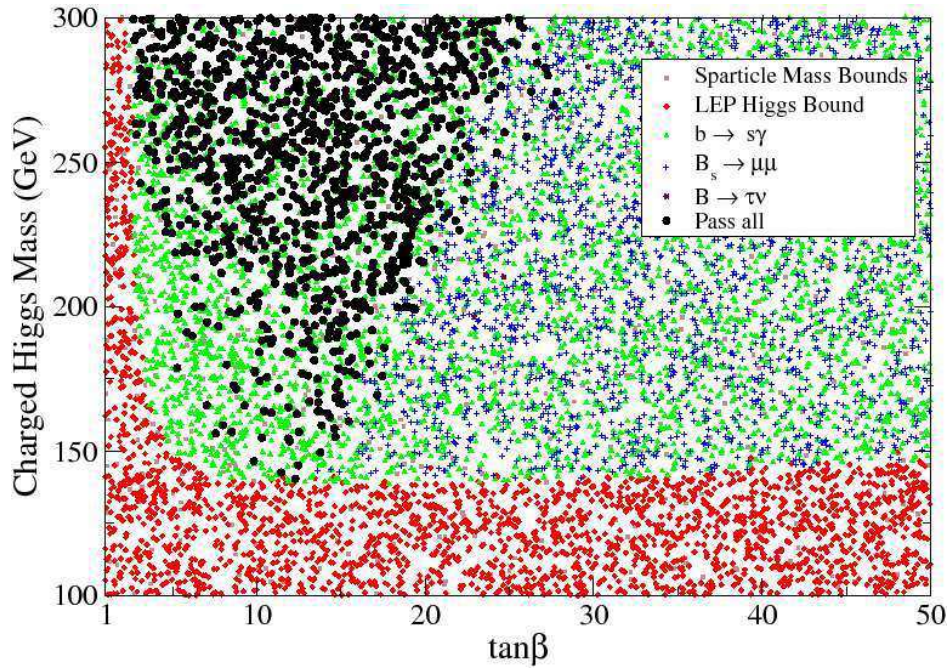}
\caption{Scan of SUSY models examined in the $\tan\beta$ vs.\ $m_{H^{\pm}}$
plane. This figure shows models of the Max-Mixing scenario. A point color/symbol indicates the first constraint which ruled it out, or is a dark black circle if it was not ruled out. The order of constraints it shown in the legend box and discussed in the text. Note that only a few points were ruled out by $B\to\tau\nu$ alone, and are hard to find in the figure.
\label{MaxmixHvsTanb}}
\end{figure}

Our primary result is shown in Fig.~\ref{MaxmixHvsTanb}, showing $m_{H^{\pm}}$
vs.\ $\tan\beta$ for a set of Max-Mixing models. The electroweak and flavor constraints are applied one at a time, in the following order: {\sl (1)} all direct search constraints on sparticles are applied as discussed at the end of $\S1$; {\sl (2)} the mass bounds on $h$ and $A$ are applied also following the discussion of $\S1$; {\sl (3)} the $b\to s\gamma$ constraint is applied following $\S2.1$; {\sl (4)} the $B_s\to\mu\mu$ constraints is applied as in $\S2.2$; and finally {\sl (5)} the $B\to\tau\nu$ constraint is applied following $\S2.3$. The points are colored according to the bound which they first fail. Those points which remain after all constraints have been applied are shown as dark black, filled circles.

Some
constraints described in the text had no noticeable effect on the
parameter space. For example, the $B\to D\tau\nu$ bound (described in $\S2.4$) only
rules out model points which are already in conflict with the LEP Higgs or $B_s\to\mu\mu$
bounds. Other constraints, such as
the neutral LSP requirement, are built into our choice of parameter ranges, as
described in the previous section.

The region that passes the constraints has several interesting
properties. Most importantly, we find that a light charged Higgs mass is possible in
the MSSM, with models possessing a charged Higgs down to $140\gev$
still able to pass all constraints.  The region which allows for a
light $H^{\pm}$ is similar to the non-universal Higgs mass (NUHM) case as examined in Ref.~\cite{Eriksson:2008cx}.

It is also interesting that the lower bound on $\mhpm$ appear to come primarily from the LEP constraints on the the neutral Higgs
masses (particularly $\mh$, though it is not obvious in the figure).
In other words, it is possible to find ranges of SUSY sparticle masses and
parameters for which all the indirect observables are consistent with experiment,
for charged Higgs masses right down to the direct bounds coming from LEP. But
once the charged Higgs mass falls below $140\gev$ it appears to be impossible to
find a sets of parameters that remain consistent with the bound on $\mh$. This is a
somewhat unfortunate circumstance, since the calculation of $\mh$ is so complicated
that different calculations of $\mh$ could potentially lead to very different
bounds on $\mhpm$; this is something one of us is exploring further~\cite{Dudleythesis}.

One can also see the effect of the $b\to s\gamma$ constraint starkly. In the region of $3\lsim \tan\beta\lsim 15$, there are very few model points which are consistent with all bounds yet yield a light $H^\pm$. And as we push to lower and lower $\mhpm$, the number of points rapidly decreases. This is because such light $H^\pm$ require large cancellations in the $b\to s\gamma$ calculation, cancellations that become harder and harder to arrange the more $\mhpm$ decreases.

\begin{figure}[t]
\centering
\includegraphics[scale = .50]{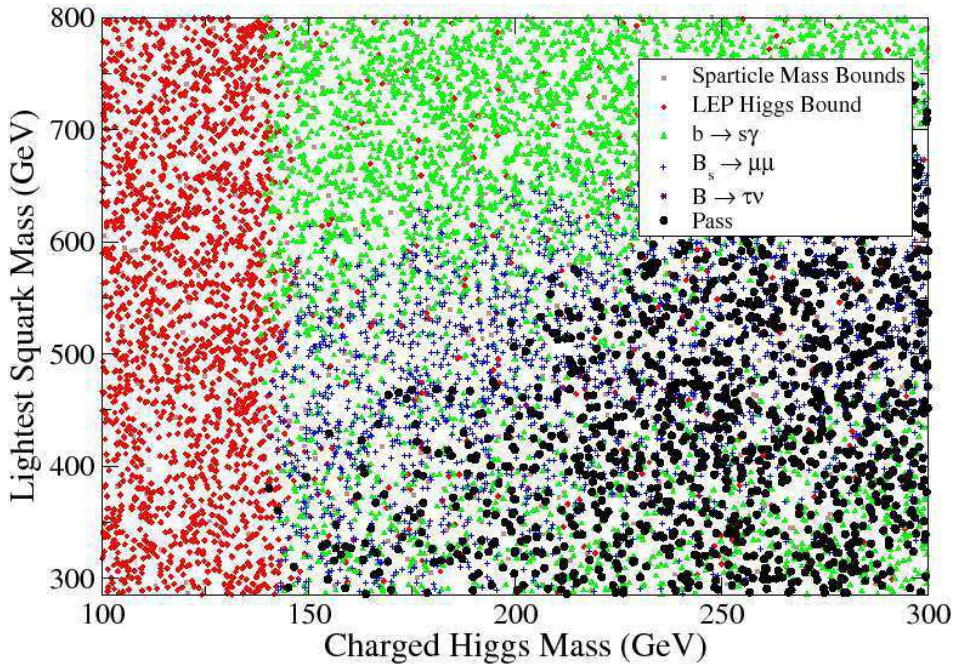}
\caption{Scan of the same MSSM models from previous figure now
comparing $m_{H^{\pm}}$ to the lightest squark mass. Models still
fall under the Max-Mixing scenario.\label{MaxmixMsqvsMhpm}}
\end{figure}

In order to arrange such a cancellation we are forced to
have other light sparticles, particularly top squarks. In Figure~\ref{MaxmixMsqvsMhpm} we
can see this explicitly. As $\mhpm$ decreases, the upper bound on the lightest squark (almost always a stop) also decreases. For $\mhpm< 200\gev$, a squark below $500\gev$ is required; for $\mhpm<155\gev$, a squark below $400\gev$ is required. This can be contrasted with the $\mhpm\simeq 300\gev$, when squark masses up to $750\gev$ are fine. Thus a light charged Higgs necessarily implies other ``light" sparticles, in particular, top squarks.

The lightest charged Higgs masses in Fig.~\ref{MaxmixHvsTanb} occur
in the region of $7\lsim\tan\beta\lsim 15$. Larger values of $\tan\beta$ are in
conflict with $B_s\to\mu\mu$, whose rate grows as
$\tan^6\beta$. This constraint is further strengthened at small
$m_{H^{\pm}}$ because the rate also scales as $1/m_A^4$ (with $m_A$ in turn scaling with $\mhpm$). Therefore, $B_s\to\mu\mu$ rules out a
large fraction of the low $\mhpm$, high $\tan\beta$
region of parameter space. (The constraint from $B\to\tau\nu$ also plays a role in this
region of parameter space, but only a few model points are ruled out by $B\to\tau\nu$ which
are not otherwise ruled out by $B_s\to\mu\mu$; these are very difficult to see in Fig.~\ref{MaxmixHvsTanb}.)
Also, in Fig.~\ref{MaxmixHvsTanb}, $\tan\beta > 50$ is not included.  This region is dominated by
$B_s\to\mu\mu$ and contains no new interesting regions.

The decay rate for $B_s\to\mu\mu$
depends on several other ingredients, an important one of which is
the parameter $A_t$. Within the Max-Mixing scenario, the value of $A_t$ is fixed
from the squark masses and is thus sizable. But in the No-Mixing scenario, $A_t$ is
set to be equal to $\mu\cot\beta$, which is very small at large $\tan\beta$. Thus it is
possible that the No-Mixing Scenario, lacking a strong $B_s\to\mu\mu$ constraint, might
allow for light $H^\pm$ over a wider range of parameters.

\begin{figure}[t]
\centering
\includegraphics[scale = .50]{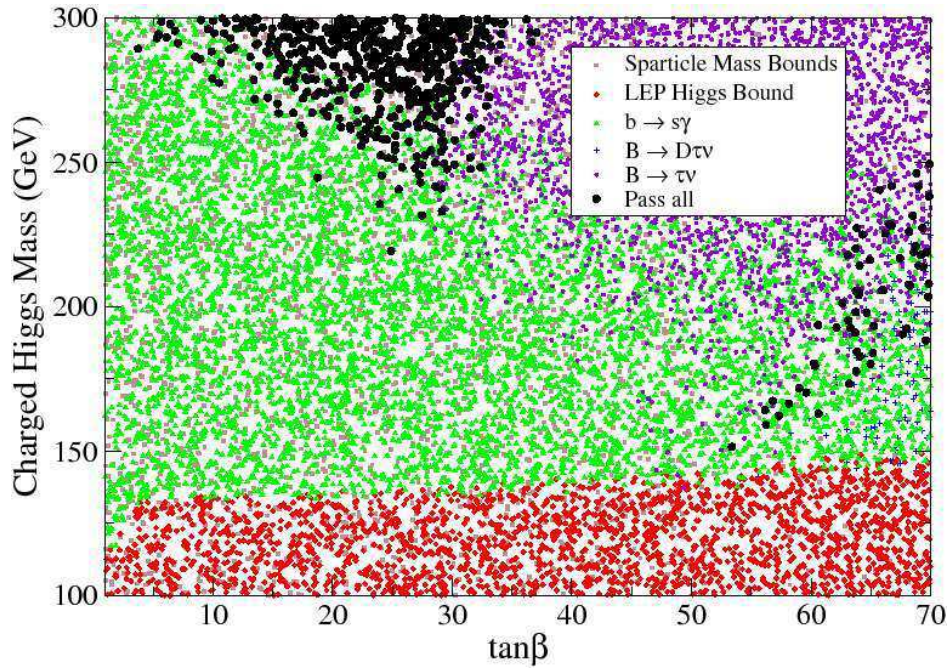}
\caption{Scan of SUSY models examined on the $\tan\beta$ vs.\ $m_{H^{\pm}}$
plane. This plot shows models which fall under the No-Mixing
scenario.\label{NomixHvsTanb}}
\end{figure}

In fact, the No-Mixing scenario changes the picture dramatically, but not as one might
expect from the arguments in the previous paragraph. Our result is shown
in Fig.~\ref{NomixHvsTanb}. As expected, the $B_s\to\mu\mu$ constraint is now entirely
absent. In its place is an even stronger $b\to s\gamma$ constraint,
and at $\tan\beta\gsim 30$, the $B\to\tau\nu$ constraint. The origin of the
strengthened $b\to s\gamma$ appears to be as follows: The No-Mixing scenario gets
its name from the absence of left-right mixing among the top squarks. But left-right
stop mixing plays a key role in several of the diagrams that help to cancel
the $H^\pm$ contribution to $b\to s\gamma$. (Recall that $b\to s\gamma$ has the form of
a magnetic moment operator and is thus chirality changing. For many parts of the
calculation, the $A$-terms provide the chirality flip.) This $b\to s\gamma$ constraint
prohibits any light charged Higgs masses from appearing at low $\tan\beta$.

Recall that in the Max-Mixing scenario, the constraint from $B\to\tau\nu$ was almost unnecessary
as the $B_s\to\mu\mu$ constraint seemed to rule out almost all models which would violate
experimental bounds on $B\to\tau\nu$. For the No-Mixing scenario, the $B_s\to\mu\mu$ constraint
is irrelevant, and the $B\to\tau\nu$ constraint plays a more interesting role.
Because the $W^{\pm}$ and $H^{\pm}$ contributions to $B\to\tau\nu$ interfere, the
high $\tan\beta$, low $m_{H^{\pm}}$ region is not completely ruled out by $B\to\tau\nu$. At very high $\tan\beta$, the charged Higgs contribution can be twice as large, and with opposite sign, as the SM piece. In this region, the $B\to\tau\nu$ constraint breaks down and a low charged Higgs mass is not
ruled out.  In our analysis, this occurs at $50 < \tan\beta < 60$, where a charged Higgs mass of about
$150\gev$ is allowed. Of course, at even larger $\tan\beta$, the charged Higgs contribution 
becomes altogether too large, ruling out
all the available low $m_{H^{\pm}}$ parameter space.

But there is one caveat: recall from our discussion in $\S2.3$ that the $B\to\tau\nu$ calculation is plagued
by uncertainties, the largest coming from the poorly measured value of $V_{ub}$ and, to a slightly lesser degree, $f_B$. 
The uncertainty in these inputs translates into an uncertainty in the boundary of the region ruled out by $B\to\tau\nu$. Depending on the precise values used, the region at $\tan\beta\gsim 55$ where we find light charged Higgs masses may grow (\ie, a light charged Higgs could be consistent even for slightly lower $\tan\beta$) or it could vanish completely (due to the competing constraint from $B\to D\tau\nu$). One
of us is examining this issue further~\cite{Dudleythesis}.

While the two main scenarios detailed above cover a important, and often studied, portions of
the low-energy MSSM parameter space, there are still large portions
of the MSSM not fully examined.  We have also examined
some other special regions of the parameter space, but found no results inconsistent with those shown above. In particular, we have dropped the constraint on the sign of $\mu$ coming from $a_\mu$; we find that cancellations in $b\to s\gamma$ are even more difficult to arrange, and so a light $H^\pm$ is even more unlikely.

\section{Conclusion}

A discovery of the charged Higgs, predicted in both trivial (2HDSM) and non-trivial (MSSM) extensions to
the Standard Model, would be an unambiguous sign of new
physics. Such a particle could be discovered directly at the LHC or through its
indirect effects on rare flavor-changing processes. In 2HDSMs in particular, the measurement
of $b\to s\gamma$ puts a severe constraint on the mass
of a charged Higgs.  However, in the MSSM, new contributions from other SUSY particles may partially cancel the charged Higgs contribution, which allows for a considerable reduction in this bound. Precise knowledge of how light the
charged Higgs is allowed to be in the MSSM is an important question for experimental searches, both direct and indirect. Several analyses have appeared over the last year examining exactly this question~\cite{Domingo:2007dx,Barenboim:2007sk,Eriksson:2008cx}. But because the parameter space of the MSSM is so vast, each has make a different set of assumptions about the sparticle mass spectrum, mostly from a top-down approach.
 
In this paper we re-examined this issue in an entirely bottom-up way. The points in parameter space which we have studied were chosen based on low-energy criteria alone, and not on their ability to unify at some ultraviolet scale. While one criticism of this paper would be our lack of unification, we feel that this is offset by our ability to find Higgs bounds which are fairly model independent. Likewise we have chosen to ignore questions of dark matter relic densities because the calculation of said densities would require knowing details of the spectrum (such as the slepton masses) which have no immediate effect on the charged Higgs mass, and which we therefore consider to be tunable. We suffice to demand that the LSP be a neutralino, which will at least afford the model a potentially viable dark matter candidate.

In order to make a bottom-up analysis simple yet useful, we made one further simplification: we examined the MSSM in the two extremes of No-Mixing and Maximal-Mixing, where each is defined by the effect of stop mixing corrections on the light scalar Higgs mass. These scenarios are frequently examined in the phenomenological SUSY literature and so provide good starting points for any more complete discussion.

The main results of this paper, in each of these two limits, can be found in Fig.~\ref{MaxmixHvsTanb} (Max-Mixing)
and Fig.~\ref{NomixHvsTanb} (No-Mixing).  For the Max-Mixing
case, the charged Higgs can be found to be as light
as $140\gev$, with a hard lower limit coming from the LEP searches for the light Higgs. This agrees well with the CMSSM analysis of Ref.~\cite{Eriksson:2008cx}, and is fairly similar
to the large, positive $A_t$ case in Ref.~\cite{Domingo:2007dx}. 
Because the Max-Mixing scenario requires relatively large $A_t$, 
the process $B_s\to\mu\mu$ rules out most of the available parameter space at high $\tan\beta$. At low to moderate $\tan\beta$, it is possible to find a charged Higgs as light as $140\gev$, but we are required to have a light set of squarks to cancel out the charged Higgs contributions to
$b\to s\gamma$. This can be seen in Fig.~\ref{MaxmixMsqvsMhpm}, where
a charged Higgs mass below $150\gev$ must be offset by squark(s) with mass below $400\gev$.

For the No-Mixing scenario, the story is quite different.
First, the constraint from $b\to s\gamma$ becomes even more powerful. This appears to be because, with small $A_t$, many of the chargino-stop diagrams become significantly weaker, as they require left-right stop mixing. Second, the $B_s\to\mu\mu$ constraints disappear entirely, as the branching ratio is proportional to $A_t^2$. But the $B_s\to\mu\mu$ constraint is mostly replaced by the $B\to\tau\nu$ constraint, which now plays a large and more complicated role, since the $W^\pm$ and $H^\pm$ contributions can cancel under some conditions.

Thus in the No-Mixing scenario at low to moderate $\tan\beta$, the strengthened $b\to s\gamma$ constraint pushed the lightest charged Higgs mass up to roughly $250\gev$. For
$\tan\beta > 50$, we found that a charged Higgs mass as low as $150\gev$ is possible, which loosely agrees
with the ``NUHM" case found in Ref.~\cite{Barenboim:2007sk}, though our choice of theoretical and experimental inputs differs from theirs in detail.

Either way, we have shown that in the oft-studied ``Max-Mixing" and ``No-Mixing" SUSY Higgs scenarios, it is possible for a charged Higgs mass as light as 140 to $150\gev$ to be observed. However such an observation requires cancellations in $b\to s\gamma$ (and perhaps other processes) that can only occur if one or both of the top squarks are lighter than 400 to $500\gev$. Thus, within the context of the MSSM, observation of such a light charged Higgs would appear to guarantee a host of other sparticle discoveries, making for a very exciting period at the LHC.

\section{Acknowledgements}
This work was partially supported by the National Science Foundation under grant PHY-0355066 and by the Notre Dame Center for Applied Mathematics.

\end{document}